\documentstyle[12pt]{article}
\begin{document}
\thispagestyle{empty}
\begin{center}
{\Large  Exact Self-Consistent Solutions to the \\ Interacting Spinor and 
              Scalar Field Equations \\ in Bianchi Type-I Space-Time }
\vskip 5mm
{\bf R.Alvarado$^{\star}$, Yu.P.Rybakov$^{\star}$, 
B.Saha$^{\dagger}$, G.N.Shikin$^{\star}$.} \\ 
$\star$ Russian Peoples' Friendship University, Moscow, \\ 
$\dagger$ Joint Institute for Nuclear Research, Dubna, Moscow region.  
\vskip 5mm
Summary                             
\end{center}
        
{\it   Self-consistent solutions to the system of spinor and scalar field  
equations in General Relativity are studied for the case of Bianchi  
type-I space-time. It  should  be  emphasized  the  absence  of  initial 
singularity for some types of solutions and  also  the  isotropic 
mode of space-time expansion in some special cases.}
\newpage
   The aim of the paper is to  find  some  exact  self-consistent 
solutions to the equations, describing spinor and scalar field system with 
the interaction Lagrangian 
$L_{int}=\varphi_{,\alpha}\varphi^{,\alpha}\Phi(S)$, $\Phi(S)$ 
being arbitrary function of the invariant $ S=\bar \psi \psi$ for 
Bianchi type-I space-time.  Equations for partial choice of $\Phi(S)$, 
while $[\Phi(S)]^{-1}= 1+\lambda S^n$, $\lambda$ being the coupling 
constant, $n$ being some constant, have been thoroughly studied. It is 
shown that the equations, mentioned, can possess initially regular, as 
well as singular solutions, depending on the sign of $\lambda$, 
nevertheless singularity remains absent for solutions describing 
the field system with broken dominant energy condition.

The Lagrangian for the interacting system of spinor, scalar and 
gravitation fields can be written as:
\begin{equation}
L=\frac{R}{2\kappa}+\frac{i}{2} \biggl[ \bar \psi \gamma^{\mu} 
\nabla_{\mu} \psi- \nabla_{\mu} \bar \psi \gamma^{\mu} \psi \biggr] - 
m\bar \psi \psi + \frac{1}{2} \varphi_{,\alpha} \varphi^{,\alpha} \Phi(S),
\end{equation}
with $R$ being the scalar curvature, $\kappa$ being  the  Einstein's 
gravitational constant. Function $\Phi(S)=1+\lambda F(S), S= \bar \psi 
\psi,$ describes the interaction between spinor and scalar fields, 
$\lambda$ being the interaction parameter. For $\lambda=0$ the 
interaction vanishes and $\Phi(S)=1$. In this case we have 
the system of fields with minimal coupling. 

Bianchi type-I space-time metric can be chosen in the form [1]
\begin{equation} ds^2 = 
 dt^2 - a^2(t)dx^2 - b^2(t)dy^2 - c^2(t)dz^2. \end{equation}

From Lagrangian (1) we will get Einstein 
equations, spinor and scalar field equations and components of their
energy-momentum tensor. We will use Einstein 
equations for $a(t), b(t)$ and $c(t)$ in the form [1]:
\begin{eqnarray}
\frac{\ddot a}{a} +\frac{\dot a}{a} \biggl(\frac{\dot b}{b}+\frac{\dot 
c}{c}\biggr)= -\kappa \biggl(T_{1}^{1}- \frac{1}{2}T\biggr), \\
\frac{\ddot b}{b} +\frac{\dot b}{b} \biggl(\frac{\dot a}{a}+\frac{\dot 
c}{c}\biggr)= -\kappa \biggl(T_{2}^{2}- \frac{1}{2}T\biggr),  \\
\frac{\ddot c}{c} +\frac{\dot c}{c} \biggl(\frac{\dot a}{a}+\frac{\dot 
b}{b}\biggr)= -\kappa \biggl(T_{3}^{3}- \frac{1}{2}T\biggr),   \\
\frac{\ddot a}{a} +\frac{\ddot b}{b} +\frac{\ddot 
c}{c}= -\kappa \biggl(T_{0}^{0}- \frac{1}{2}T\biggr),
\end{eqnarray}
where point means differentiation with respect to t, and 
$T=T_{\mu}^{\mu}.$ 

Spinor and scalar field equations and components of 
its energy-momentum tensor can be written as follows: 
\begin{equation}
i\gamma^\mu \nabla_\mu \psi -m\psi + \frac{1}{2} \varphi_{,\alpha} 
\varphi^{,\alpha} \Phi^{\prime} (S) \psi =0, \quad \Phi^{\prime} (S)= 
\frac{d\Phi}{dS}, \end{equation} \begin{equation} 
\frac{1}{\sqrt{-g}}\frac{\partial}{\partial 
x^\nu}\biggl(\sqrt{-g}g^{\nu\mu}\varphi_{,\mu}\Phi(S)\biggl)=0, 
\end{equation}
\begin{equation}
T_{\mu}^{\rho}=\frac{i}{4} g^{\rho\nu} \biggl(\bar \psi \gamma_\mu 
\nabla_\nu \psi + \bar \psi \gamma_\nu \nabla_\mu \psi - \nabla_\mu \bar 
\psi \gamma_\nu \psi - \nabla_\nu \bar \psi \gamma_\mu \psi \biggr) + 
\varphi_{,\mu} \varphi^{,\rho} \Phi(S) - \delta_{\mu}^{\rho}L.
\end{equation} 
In (7) and (9) $\nabla_\mu$ denotes covariant derivative of spinor, having 
the form [2]:  \begin{equation} \nabla_\mu \psi=\frac{\partial 
\psi}{\partial x^\mu} -\Gamma_\mu \psi, \end{equation}
where $\Gamma_\mu(x)$ are spinor affine connection matrices.  
$\gamma^\mu(x)$ matrices are defined for the metric (2) 
as follows. Using the equality
$$ g_{\mu \nu} (x)= e_{\mu}^{a}(x) e_{\nu}^{b}(x) \eta_{ab}, \quad
\gamma_\mu(x)= e_{\mu}^{a}(x) \bar \gamma_a,$$
where $\eta_{ab}= diag(1,-1,-1,-1)$, $\bar \gamma_a$ being flat 
space-time Dirac matrices, $e_{\mu}^{a}$ denoting a set of tetrad 
4-vectors, we will get $$ \gamma^0=\bar \gamma^0,\quad \gamma^1 =\bar 
\gamma^1 /a(t),\quad \gamma^2= \bar \gamma^2 /b(t),\quad \gamma^3 = \bar 
\gamma^3 /c(t). $$ $\Gamma_\mu(x)$ matrices are defined by the equality 
$$\Gamma_\mu (x)= 
\frac{1}{4}g_{\rho\sigma}(x)\biggl(\partial_\mu e_{\delta}^{b}e_{b}^{\rho} 
- \Gamma_{\mu\delta}^{\rho}\biggr)\gamma^\sigma\gamma^\delta, $$ 
which gives
\begin{equation} \Gamma_0=0, \quad \Gamma_1=\frac{1}{2}\dot a(t) 
\bar \gamma^1 \bar \gamma^0, \quad \Gamma_2=\frac{1}{2}\dot b(t) \bar 
\gamma^2 \bar \gamma^0, \quad \Gamma_3=\frac{1}{2}\dot c(t) \bar \gamma^3 
\bar \gamma^0,\end{equation}
Flat space-time matrices we will choose in the form, given in [3]. 
 
We will study the space-independent solutions to spinor and scalar field 
equations (7), (8) so that
$$\psi=V(t), \quad \varphi =\varphi(t).$$
In this case solution to the equation (8) is:
\begin{equation}
\dot \varphi(t)=\frac{C}{\tau \Phi(S)}, \quad C=const, \quad 
\tau(t)=a(t)b(t)c(t). \end{equation} 
In accordance with (12) the spinor field equation (7) can be written as:
\begin{equation} i\bar \gamma^0 
\biggl(\frac{\partial}{\partial t} +\frac{\dot \tau}{2 \tau} \biggr)V -mV 
-\frac{C^2}{2 \tau^2}P^{\prime} (S) V=0, \end{equation} 
where $ P(S)=1/\Phi(S);\quad P^{\prime}(S)= \frac{dP}{dS}= 
-\frac{\Phi^{ \prime}}{\Phi^2}.$ For the components 
$\psi_\rho= V_\rho(t), \quad \rho=1,2,3,4,$ from (13) one deduces
the following system of equations:  
\begin{eqnarray} \dot V_r +\frac{\dot \tau}{2 \tau} V_r 
+i\biggl(m+\frac{C^2 P^{\prime}}{2 \tau^2}\biggr) V_r &=& 0, \quad r=1,2; 
\\ \dot V_l +\frac{\dot \tau}{2 \tau} V_l -i\biggl(m+\frac{C^2 
P^{\prime}}{2 \tau^2}\biggr) V_l &=& 0, \quad l=3,4.  \end{eqnarray} 

From (14) and (15) we will find  the  equation for invariant function 
$$S=\bar \psi \psi= V_{1}^{\ast} V_1 + V_{2}^{\ast} V_2 - V_{3}^{\ast} V_3 - 
V_{4}^{\ast} V_4:$$  
\begin{equation} \dot S +\frac{\dot \tau}{\tau}S=0, 
\end{equation}
which leads to
\begin{equation}
S=\frac{C_0}{\tau}, \quad C_0= const.
\end{equation}

As in the considered case $P$ depends only on  S, from  (17) it
follows that $P(S)$ and $P^{\prime}(S)$ are functions of $\tau= abc$. 
Taking this fact into account, integration  of the system of equations 
(14) and (15) leads to the expressions 
\begin{eqnarray}
V_{r}(t) &=& \frac{C_{r}}{\sqrt{\tau}}exp\biggl[-i(mt+ \int Q 
dt)\biggr], \qquad r=1,2; \nonumber \\
V_{l}(t) &=& \frac{C_{l}}{\sqrt{\tau}}exp\biggl[i(mt+ \int Q 
dt)\biggr], \qquad l=3,4. 
\end{eqnarray}
where $Q(t)= \frac{C^2 P^{\prime}}{2 \tau^2}, \quad C_r$ ¨ $C_l$ - 
integration constants.

Putting (18) into (9), we will get the following expressions for the 
components of the energy-momentum tensor for the interacting spinor 
and scalar fields  

\begin{equation}
T_{0}^{0}=\frac{i}{2}N + \frac{C^2}{\tau^2}P- R, \quad 
T_{1}^{1}=T_{2}^{2}=T_{3}^{3}=-R, \end{equation}
where
\begin{equation}
N=-\frac{2i}{\tau}(C_{1}^{2}+C_{2}^{2}-C_{3}^{2}-C_{4}^{2})(m+Q) = 
-\frac{2iC_0}{\tau}\biggl(m+\frac{C^2 P^{\prime}}{2 \tau^2}\biggr); 
\end{equation} 
\begin{equation}
R=\frac{C^2}{2 \tau^2}\biggl(P+\frac{C_0 P^{\prime}}{\tau}\biggr), \quad 
T=T_{\alpha}^{\alpha}=\frac{i}{2}N + \frac{C^2}{\tau^2}P- 4R.
\end{equation}
Summation of Einstein equations (3),(4) and (5) leads  to the equation 
\begin{equation}
\frac{\ddot \tau}{\tau}= -\kappa \biggl(T_{1}^{1}+T_{2}^{2}+T_{3}^{3}- 
\frac{3}{2}T\biggr)= 3\kappa \biggl(\frac{mC_0}{2 \tau}-\frac{C_0 C^2 
P^{\prime}}{4 \tau^3}\biggr). \end{equation}
The first integral of the equation (22) takes the form:
\begin{equation}
\dot \tau^2 = 3\kappa (mC_0 \tau +\frac{1}{2}C^2 P +C_1), \quad C_1= 
const. \end{equation}
Final solution of the equation (22) reads
\begin{equation}
\int \frac{d \tau}{\sqrt{(mC_0 \tau +\frac{1}{2}C^2 P +C_1)}}= \pm 
\sqrt{3\kappa}(t+ t_0), \quad t_0=const. \end{equation}
Giving the explicit form of $\Phi(S)$, i.e. $P=1/\Phi$, from 
(24) one  can find concrete function $\tau(t)=a(t)b(t)c(t)$. Putting the 
obtained function in (18), one can get expressions for components of 
spinor function $V_{\rho}(t)$, where $\rho =1,2,3,4.$

Let us express $a, b, c$ through $\tau$. For this we notice that
subtraction of Einstein equations  (3)-(4)  leads  to  
the equation 
\begin{equation}
\frac{\ddot a}{a}-\frac{\ddot b}{b}+\frac{\dot a \dot c}{ac}- 
\frac{\dot b \dot c}{bc}= \frac{d}{dt}\biggl(\frac{\dot a}{a}- 
\frac{\dot b}{b}\biggr)+\biggl(\frac{\dot a}{a}- \frac{\dot b}{b} \biggr) 
\biggl (\frac{\dot a}{a}+\frac{\dot b}{b}+ \frac{\dot c}{c}\biggr)= 0. 
\end{equation} 
Equation (25) possesses the solution
\begin{equation}
\frac{a}{b}= D_1 exp \biggl(X_1 \int \frac{dt}{\tau}\biggr), \quad 
D_1=const, \quad X_1= const. \end{equation}
Subtracting equations (3)-(5) and (4)-(5) one finds the 
equations similar to (25), having solutions   
\begin{equation} 
\frac{a}{c}= D_2 exp \biggl(X_2 \int \frac{dt}{\tau}\biggr), \quad 
\frac{b}{c}= D_3 exp \biggl(X_3 \int \frac{dt}{\tau}\biggr),  
\end{equation}
where $D_2, D_3, X_2, X_3 $ are integration  constants. There is a 
functional dependence between the constants 
$D_1, D_2, D_3, X_1, X_2, X_3 $:  
\begin{equation}
D_2=D_1 D_3, \qquad X_2= X_1+X_3.
\end{equation}
Using the equations (26), (27) ¨ (28), we rewrite $a(t), b(t), c(t)$ in 
the explicit form:  
\begin{eqnarray} a(t) &=& 
(D_{1}^{2}D_{3})^{\frac{1}{3}}\tau^{\frac{1}{3}}exp\biggl[\frac{2X_1 
+X_3}{3} \int \limits_{t_0}^{t}\tau^{-1}dt' \biggr] \nonumber \\
b(t) &=& 
(D_{1}^{-1}D_{3})^{\frac{1}{3}}\tau^{\frac{1}{3}}exp\biggl[-\frac{X_1 
-X_3}{3} \int \limits_{t_0}^{t}\tau^{-1}dt' \biggr] \nonumber \\
c(t) &=& 
(D_{1}D_{3}^{2})^{-\frac{1}{3}}\tau^{\frac{1}{3}}exp\biggl[-\frac{X_1 
+2X_3}{3} \int \limits_{t_0}^{t}\tau^{-1}dt' \biggr] 
\end{eqnarray}
where $t_0$ is the initial time.

Thus the previous system of Einstein equations  and interacting spinor and 
scalar field ones is completely integrated. In  this  process of 
integration only first three of the complete system  of  Einstein 
equations have been used.  General solutions to these three second order 
equations have been obtained. The  solutions  contain  six arbitrary 
constants: $D_1, D_3, X_1, X_3 $  and two others $C_1$ and $t_0$, 
that were obtained while  
solving  equation  (22).  Equation  (6)  is  the consequence of first 
three of Einstein equations.  To  verify  the correctness of obtained 
solutions, it is necessary to put $a(t), b(t)$ and $c(t)$ in (6). It 
should lead either to identity or to  some  additional constraint between 
the constants.  Putting $a(t), b(t), c(t)$ from (29) in (6) one can get 
the following equality:
\begin{equation}
\frac{1}{3 \tau}\biggl[3 \ddot \tau - 2\frac{\dot \tau^2}{\tau}+ 
\frac{2}{3 \tau}\biggl(X_{1}^{2}+X_1 X_3 +X_{3}^{2}\biggr)\biggr] = 
-\kappa \biggl(T_{0}^{0}-\frac{1}{2}T\biggr), 
\end{equation}
that guaranties the correctness of obtained solutions.  

To get the constant $C_1$ in (24) one can use the equation (30). 
Inserting $\ddot \tau$  from (22), $\dot \tau^2$ from (23) and
$$ T_{0}^{0}- \frac{1}{2}T= 
\frac{i}{4}N+\frac{1}{2}\frac{C^2}{\tau^2}P+R =\frac{mC_0}{2 \tau}+ 
\frac{C^2}{\tau^2}P+\frac{3C_0 C^2}{4 \tau^3}P^{\prime},$$ one deduces
the identity, if 
\begin{equation} C_1=\frac{1}{9 
\kappa} \biggl(X_{1}^{2}+X_1 X_3 +X_{3}^{2}\biggr), \end{equation} 
which means that the constant $C_1$ is a positive one.

We will first study the solution to the system of field equations with 
minimal coupling when the direct interaction between the spinor and 
scalar fields remains absent, i.e. in the Lagrangian (1) $\Phi(S) \equiv 
1$. The reason to get the solution  to the self-consistent system of 
equations for the fields with minimal coupling is the necessity 
of comparing this solution with that for the  system  of equations   for 
the interacting spinor, scalar and gravitational fields  that permits  to 
clarify the  role of interaction terms in the evolution of the 
cosmological model in question. 

In this case the components of the energy-momentum tensor look:
\begin{eqnarray}
T_{0}^{0} &=& \frac{mC_0}{\tau}+ \frac{C^2}{2 \tau^2}, \quad 
T_{1}^{1}=T_{2}^{2}=T_{3}^{3}=- \frac{C^2}{2 \tau^2},  \nonumber \\
T &=& T_{\alpha}^{\alpha}=\frac{mC_0}{\tau}- \frac{C^2}{\tau^2}, \quad
T_{1}^{1}+T_{2}^{2}+T_{3}^{3}-\frac{3}{2}T =-\frac{3}{2}\frac{mC_0}{\tau}.
\end{eqnarray}
Note that as the energy density $T_{0}^{0}$ should be a quantity 
positively defined, the equation (32) leads to $C_0 >0$. The inequality 
$C_0 >0$ will also be preserved for the system with direct interaction 
between the fields as in this case the correspondence principle should be 
fulfilled: for $\lambda =0$ the field system with direct interaction turns 
into that with minimal coupling.

Taking into account (32) equation (22) writes 
\begin{equation}
\ddot \tau = \frac{3}{2}\kappa m C_0,
\end{equation}
with the solution
\begin{equation}
\tau(t) = \frac{3}{4}\kappa m C_0 t^2 + \tau_1 t + \tau_2, \quad \tau_1, 
\tau_2 = const.
\end{equation}

Putting $\tau(t)$ from (34) into (18) and (29) one gets the explicit 
expressions for the components of spinor field functions $V_{\rho}(t)$ and 
metric functions $a(t), b(t), c(t)$:  \begin{equation} V_{r}(t) = 
\frac{C_{r}}{\sqrt{\tau}}e^{-imt}, \quad V_{l}(t) = 
\frac{C_{l}}{\sqrt{\tau}}e^{imt}, \end{equation} \begin{eqnarray} a(t) &=& 
(D_{1}^{2}D_{3})^{\frac{1}{3}}\tau^{\frac{1}{3}}Z^{\frac{2X_1 
+X_3}{3}}, \nonumber \\
b(t) &=& 
(D_{1}^{-1}D_{3})^{\frac{1}{3}}\tau^{\frac{1}{3}}Z^{-\frac{X_1 
-X_3}{3}}, \nonumber \\
c(t) &=& 
(D_{1}D_{3}^{2})^{-\frac{1}{3}}\tau^{\frac{1}{3}}Z^{-\frac{X_1 
+2X_3}{3}}, 
\end{eqnarray}
where 
\begin{equation}
Z=\Biggl(\frac{t-t_1}{t-t_2}\Biggr)^\sigma, \quad 
\sigma=\frac{4}{3\kappa m C_0 (t_1 -t_2)}, 
\end{equation}
and $ t_{1,2}= -\frac{2 \tau_1}{3\kappa m C_0} \pm \frac{2}{3\kappa m 
C_0}\sqrt{\tau_{1}^{2}- 3\kappa m C_0 \tau_2}$  are the roots of the 
quadratic polinomial in the right-hand side of (34). If 
the roots are real, i.e. if \begin{equation} \tau_{1}^{2}- 3\kappa m C_0 
\tau_2 \geq 0, \end{equation} the solution (34) is singular one, while in 
the opposite case it is not.  Putting (34) into (30) one deduces the 
following relation between the constants:  \begin{equation} \tau_{1}^{2}- 
3\kappa m C_0 \tau_2 = \frac{3}{2}\kappa C^2 + \frac{1}{3} 
\biggl(X_{1}^{2}+X_1 X_3 +X_{3}^{2}\biggr). 
\end{equation}
As the right-hand side of the equation (39) is positive, the quadratic 
trinomial in (34) possesses real roots and the solution obtained is 
singular one at initial time $t=t_1$, whereas $t_1 > t_2, t_1 \leq t \leq 
\infty.$

Let us study the solution (34)-(36) at $t \to \infty$. Hence we have: 
\qquad $\tau (t) \approx \frac{3}{4}\kappa m C_0 t^2,$  \quad and $a(t) 
\sim b(t) \sim c(t) \sim t^{2/3},$ that leads to the conclusion about the 
asymptotical isotropization  of  the expansion process for the initially 
anisotropic Bianchi  type-I space-time.  

Thus the solution to the self-consistent system  of  equations for the 
spinor, scalar and 
gravitational fields  is  the  singular one at the initial time. In the 
initial state of evolution of the field system the expansion process of 
space-time is  anisotropic,  but at $t \to \infty$ there happens 
isotropization of the expansion process.    

To investigate the system of spinor and scalar field equations with direct 
interaction we will consider the partial case for choosing $P(S)$:
\begin{equation}
P(S)= 1+ \lambda S^n = 1 + \lambda \frac{C_{0}^{n}}{\tau^n},
\end{equation}
where $\lambda$ is the interaction parameter, n is some arbitrary 
constant. Inserting (40) into (24) one obtains 
\begin{equation} \int \frac{d \tau}{\sqrt{mC_0 \tau +\lambda C^2 
C_{0}^{n}/2 \tau^n + C_{2}^{2}}}= \sqrt{3\kappa}t, \end{equation} where 
$C_{2}^{2}= C^2/2+C_1$; in (24) $t_0$ has been taken zero, as it only 
gives the shift of the initial time. 

Let us study different cases of choosing $\lambda$ and n.  
I. $\lambda > 0, \quad n>0.$ In this case (41) leads to the following 
behavior of $\tau(t)$: \begin{equation} at \quad t \to \infty \quad 
\tau(t) \approx \frac{3}{4}\kappa m C_0 t^2 \to \infty, \end{equation} 
\begin{equation} at \quad t \to 0 \quad \tau(t) \approx 
\Biggl[\biggl(\frac{n}{2}+1\biggr) \sqrt{\frac{3\kappa \lambda 
C_{0}^{n}C^2}{2}}t\Biggr]^\frac{1}{n/2 +1} \to 0.  \end{equation} Note 
that (42) coincides with (34) at $t \to \infty$. It leads to the fact that 
in the case considered, the asymptotical isotropization of the expansion 
process of initially anisotropic Bianchi type-I space-time takes place 
without the influence of scalar field. In this case the initial state is 
singular: $ \tau (0)=0.$

Thus, the evolution of the interacting fields system at
$\lambda >0$ and $n>0$ is qualitatively the same as that of the system 
with minimal coupling.

II. $\lambda = -\sigma^2 <0, \quad n>0$. The equation (41) takes the form:
\begin{equation} \int \frac{d \tau}{\sqrt{mC_0 \tau 
-\sigma^2 C^2 C_{0}^{n}/2 \tau^n + C_{2}^{2}}}= \sqrt{3\kappa}t.  
\end{equation} From (44) follows:  $$ at \quad t \to \infty 
\quad \tau(t) \approx \frac{3}{4}\kappa m C_0 t^2 \to \infty,$$ i.e. 
as well as in the previous case  the asymptotical isotropization of the 
expansion process of initially anisotropic Bianchi type-I space-time takes 
place. But $\tau =0$ cannot  be  reached as in  
this case the denominator of the integrand in (44) becomes imaginary at 
$\tau \to 0$. There exists the minimum value $\tau_{min}= \tau_0 >0$, 
which is defined from the equation
$$mC_0 \tau_{0}^{n+1}+ C_{2}^{2} 
\tau_{0}^{n}-\frac{\sigma^2 C^2 C_{0}^{n}}{2}=0.$$ 
It means that for $\lambda<0$ and $n>0$ there exist regular solutions to 
the previous system of equations. The absence  of  the  initial  
singularity in  the considered cosmological solution appears to  be  
consistent  with the violation for $\lambda<0$, of the dominant energy 
condition in   the Hawking-Penrose theorem [1].

III. $\lambda >0, \quad n=-k^2 <0$. In this case the equation (41) takes 
the form:
\begin{equation} \int \frac{d \tau}{\sqrt{mC_0 
\tau +\lambda C^2 \tau^{k^2}/2 C_{0}^{k^2} + C_{2}^{2}}}= \sqrt{3\kappa}t.  
\end{equation} 
Let us study concrete solutions for some values of $k^2$.

a) $k^2 =1$. Then from (45) one gets:
\begin{equation}
\tau(t)= \frac{3}{4}MC_0 \kappa t^2 - \frac{C_{2}^{2}}{MC_0}, \quad M=m + 
\frac{\lambda C^2}{2C_{0}^{2}}.
\end{equation} 
The solution (46) is singular one at initial time 
$t_0 = \frac{2C_2}{\sqrt{3\kappa}MC_0}$ and asymptotically isotropic.

b) $k^2 =2$. The equation (45) writes
\begin{equation}
\int \frac{d \tau}{\sqrt{mC_0 \tau +\lambda C^2 \tau^2 /2 
C_{0}^{2} + C_{2}^{2}}}= \sqrt{3\kappa}t. 
\end{equation} 
Integration of (47) leads to  
\begin{equation} \tau(t)=\frac{C_{0}^{2}}{\lambda 
C^2}\Biggl[\triangle sh \Biggl(\frac{\sqrt{3\kappa 
\lambda}C}{\sqrt{2}C_0}t\Biggr)-mC_0\Biggr], \end{equation} $$\triangle = 
\sqrt{2\lambda C^2 C_{2}^{2}/C_{0}^{2}- m^2 C_{0}^{2}}.$$ 
From (48) one gets: $\tau(t_0)=0$, where $t_0$ is defined from the 
equation:  
\begin{equation} \triangle sh 
\Biggl(\frac{\sqrt{3\kappa \lambda}C}{\sqrt{2}C_0}t_0\Biggr)-mC_0 =0, 
\end{equation}
i.e. the solution (48) is singular at initial time $t=t_0$.

At $t \to \infty$
\begin{equation}
\tau(t) \approx \frac{C_{0}^{2}}{\lambda C^2}
\triangle exp \Biggl(\frac{\sqrt{3\kappa 
\lambda}C}{\sqrt{2}C_0}t\Biggr).
\end{equation}
The solution (48) describes initially (i.e. at $t_0$) singular and 
asymptotically (i.e. at $t \to \infty$) isotropic Bianchi Type-I 
cosmological model. Note that in this case the transition to the 
isotropic regime happens exponentially.

IV. $\lambda = -\sigma^2 <0, \quad n=- k^2 <0.$  In this case the equation 
(41) takes the form:
\begin{equation}
\int \frac{d \tau}{\sqrt{mC_0 \tau -\sigma^2 C^2 \tau^{k^2} /2 
C_{0}^{k^2} + C_{2}^{2}}}= \sqrt{3\kappa}t. 
\end{equation} 
Let us consider concrete solutions for some values of $k^2$ as in III. 

a) $k^2=\frac{1}{2}$. In this case one gets from (51):
\begin{equation}
\frac{2}{\sqrt{mC_0}}\biggl(\sqrt{\tau}+\sqrt{\tau_1}ln \mid \sqrt{\tau}- 
\sqrt{\tau_1}\mid \biggr)= \sqrt{3\kappa}t, \quad 
\sqrt{\tau_1}=\frac{\sigma^2 C^2}{4mC_{0}^{3/2}}, 
\end{equation} 
which leads to
\begin{equation}
at \quad t \to \infty \quad \tau(t) \approx \frac{3\kappa mC_0}{4}t^2 
\to \infty, \end{equation} \begin{equation} at \quad t \to -\infty 
\quad \tau(t) \to \tau_1 = \Biggl(\frac{\sigma^2 
C^2}{4mC_{0}^{3/2}}\Biggr)^2.  \end{equation} 
From (53) and (54) one comes to the conclusion that the solution (52) is 
initially (i.e. at $t_0 = -\infty$) regular one and at
$t \to \infty$ asymptotically isotropic.

b) $k^2=1$. In this case (51) writes:
\begin{equation}
\int \frac{d \tau}{\sqrt{\biggl(m -\sigma^2 C^2 /2 C_{0}^{2}\biggr)C_0 \tau + 
C_{2}^{2}}}= \sqrt{3\kappa}t. 
\end{equation} 
If in (55) $m -\sigma^2 C^2 /2 C_{0}^{2}>0,$ then the solution coincides 
with (46), where $M=m -\sigma^2 C^2 /2 C_{0}^{2}$. At $m 
-\sigma^2 C^2 /2 C_{0}^{2}=- T^2 <0$ from (55) one gets  \begin{equation} 
\tau(t)= \frac{C_{2}^{2}}{T^2 C_0}- \frac{3\kappa T^2 C_0}{4}t^2.  
\end{equation} 
In this case $\tau(t)$ possesses \begin{eqnarray} maximum 
\quad at \quad t=0, \quad i.e. \quad \tau(0)= \frac{C_{2}^{2}}{T^2 
C_0}, \nonumber \\ and \quad minimum \quad at \quad t_{1,2}=\mp 
\frac{2C_2}{\sqrt{3\kappa}T^2 C_0} \quad i.e. \quad \tau(t_{1,2})=0.  
\end{eqnarray} 
The solution obtained describes the cosmological model, which begins to 
expand at $t_1$, acquires its maximum at $t=0$ and then collapses 
into a point at $t_2$.

c) $k^2=2.$ From (45) one gets:
\begin{equation}
\int \frac{d \tau}{\sqrt{mC_0 \tau -\sigma^2 C^2 \tau^2/2 C_{0}^{2} + 
C_{2}^{2}}}= \sqrt{3\kappa}t. 
\end{equation} 
Integrating (58), for $\tau(t)$ one gets the following expression:
\begin{equation}
\tau(t)=\frac{C_{0}^{2}}{\sigma^2 C^2} \Biggl[mC_0 +\triangle sin \Biggl( 
\frac{\sigma C \sqrt{3\kappa}t}{\sqrt{2}C_0}\Biggr)\Biggr],
\end{equation}
where $\triangle = \Biggl(m^2 C_{0}^{2}+\frac{2\sigma^2 C^2 
C_{2}^{2}}{C_{0}^{2}}\Biggr)^{1/2}$. From (59) follows that 
$\tau(t_0)=0$, where \begin{equation} 
t_0=-\frac{\sqrt{2}C_0}{\sqrt{3\kappa}\sigma 
C}arcsin\biggl(\frac{mC_0}{\triangle}\biggr),
\end{equation}
then acquires maximum
$$\tau(t_{max})=\frac{C_{0}^{2}}{\sigma^2 C^2}\biggl(mC_0+ 
\triangle\biggr), $$
where
$$ t_{max}=\frac{\pi \sqrt{2}C_0}{2\sqrt{3\kappa}\sigma C},$$
and further at $t=t_1$ again turns to zero: 
$\tau(t_1)=0$, where $$t_1=\pi+\frac{\sqrt{2}C_0}{\sqrt{3\kappa}\sigma 
C} arcsin\biggl(\frac{mC_0}{\triangle}\biggr).$$
              
Thus the solution (59) describes the cosmological model, which 
begins to expand at $t_0$, acquires its maximum at $t_{max}$ and then 
collapses into a point at $t_1$.

\vskip 5mm
One of the authors (B.Saha) is grateful to the fond MNTP "Fundamental 
Metrology" grant 2.51 for financial support. 
\vskip 2cm
{\bf References}  \vskip 2mm 
1. {\it Zeldovich Ya.B., Novikov I.D.} Construction and Evolution of 
the Universe. Moscow: "Nauka", 1975.-735p. 

2. {\it Zhelnorovich V.A.} Theory of Spinors and its Application in 
Physics and Mechanics. Moscow: "Nauka", 1982.- 270p. 
 
3. {\it Bogolubov N.N., Shirkov D.V.} An Introduction to the 
Theory of Quantized Fields. Moscow: "Nauka", 1976.-479p. 
\end{document}